\documentclass[fleqn,11pt,twoside]{article}

\usepackage{amsthm,amsthm,amssymb, color, xcolor,epsfig, graphics, subfigure}

\usepackage{amsmath, graphicx, latexsym, lscape }

\makeatletter
\newcommand{\copyrightnote}[2]{{\renewcommand{\thefootnote}{}
 \footnotetext{\small\it
\begin{flushleft}
 \copyright \ #1   #2  
\end{flushleft}}}}

\newcommand{\Name}[1]{\begin{flushleft}
                       \LARGE \bf #1
                       \end{flushleft}\vspace{-3mm}}

\newcommand{\Author}[1]{\begin{flushleft}
                       \it #1 \end{flushleft}}

\newcommand{\Date}[1]{\begin{flushleft}
                      \small  \it #1 \end{flushleft}}

\newcommand{\evenhead}{Author \ name}
\newcommand{\oddhead}{Article \ name}

\renewcommand{\@evenhead}{
\hspace*{-3pt}\raisebox{-15pt}[\headheight][0pt]{\vbox{\hbox to \textwidth
{\thepage \hfil \evenhead}\vskip4pt \hrule}}}
\renewcommand{\@oddhead}{
\hspace*{-3pt}\raisebox{-15pt}[\headheight][0pt]{\vbox{\hbox to \textwidth
{\oddhead \hfil \thepage}\vskip4pt\hrule}}}
\renewcommand{\@evenfoot}{}
\renewcommand{\@oddfoot}{}

\long\def\@makecaption#1#2{%
  \vskip\abovecaptionskip
  \sbox\@tempboxa{\small \textbf{#1.}\ \ #2}%
  \ifdim \wd\@tempboxa >\hsize
    {\small \textbf{#1.}\ \ #2}\par
  \else
    \global \@minipagefalse
    \hb@xt@\hsize{\hfil\box\@tempboxa\hfil}%
  \fi
  \vskip\belowcaptionskip}

\newcommand{\JNMPnumberwithin}[3][\arabic]{%
  \@ifundefined{c@#2}{\@nocounterr{#2}}{%
    \@ifundefined{c@#3}{\@nocnterr{#3}}{%
      \@addtoreset{#2}{#3}%
      \@xp\xdef\csname the#2\endcsname{%
        \@xp\@nx\csname the#3\endcsname .\@nx#1{#2}}}}%
}

\newcommand{\resetfootnoterule} {
  \renewcommand\footnoterule{%
  \kern-3\p@
  \hrule\@width.4\columnwidth
  \kern2.6\p@}
}

\renewcommand{\footnoterule}{}

\makeatother

\theoremstyle{definition}

\begin{document}

\renewcommand{\evenhead}{ {\LARGE\textcolor{blue!10!black!40!green}{{\sf \ \ \ ]ocnmp[}}}\strut\hfill 
C Rogers and P Amster
}
\renewcommand{\oddhead}{ {\LARGE\textcolor{blue!10!black!40!green}{{\sf ]ocnmp[}}}\ \ \ \ \   
An extended modified Kadomtsov-Petviashvili equation
}

\thispagestyle{empty}
\newcommand{\FistPageHead}[3]{
\begin{flushleft}
\raisebox{8mm}[0pt][0pt]
{\footnotesize \sf
\parbox{150mm}{{Open Communications in Nonlinear Mathematical Physics}\ \  \ {\LARGE\textcolor{blue!10!black!40!green}{]ocnmp[}}
\ \ Vol.6 (2026) pp
#2\hfill {\sc #3}}}\vspace{-13mm}
\end{flushleft}}

\FistPageHead{1}{\pageref{firstpage}--\pageref{lastpage}}{ \ \ Article}

\strut\hfill

\strut\hfill

\copyrightnote{The author(s). Distributed under a Creative Commons Attribution 4.0 International License}

\Name{An Extended Modified Kadomtsov-Petviashvili Equation: Ermakov-Painlev\'e II Symmetry Reduction with Moving Boundary Application}

\Author{Colin Rogers$^1$ and Pablo Amster$^2$}

{$^1$ School of Mathematics and Statistics,}
{University of New South Wales, Sydney, Australia.}
{{\small Email: {c.rogers@unsw.edu.au}}}

\smallskip

{$^2$ Depto. de Matem\'atica, Facultad de Ciencias Exactas y Naturales,} {Universidad de Buenos Aires \& IMAS-CONICET.}
   {Ciudad Universitaria. Pabell\'on I (1428), Buenos Aires, Argentina}
{
   {\small Email: pamster@dm.uba.ar}
}
\Date{Received March 18, 2026; Accepted March 28, 2026}

\setcounter{equation}{0}

\smallskip

\noindent
{\bf Citation format for this Article:}\newline
Colin Rogers and Pablo Amster, 
An extended modified Kadomtsov-Petviashvili equation: Ermakov-Painlevé II symmetry reduction with moving boundary application,
{\it Open Commun. Nonlinear Math. Phys.}, {\bf 6}, ocnmp:17767, \pageref{firstpage}--\pageref{lastpage}, 2026.

\strut\hfill

\noindent
{\bf The permanent Digital Object Identifier (DOI) for this Article:}\newline
{\it 10.46298/ocnmp.17767}

\strut\hfill

\begin{abstract}
\noindent 
Here, a novel 2+1-dimensional nonlinear evolution equation with temporal modulation is introduced which admits integrable Ermakov-Painlev\'e II symmetry reduction. Application is made to obtain exact solution to a class of Stefan-type moving boundary problems for this 2+1-dimensional nonlinear evolution equation. Involutory transformations with origin in autonomisation of certain Ermakov-type coupled systems are extended to 2+1-dimensions and applied to derive a wide 2+1-dimensional class with temporal modulation and which inherits the property of admittance of such hybrid Ermakov-Painlev\'e II symmetry reduction applicable to certain moving boundary problems.
\end{abstract}

\label{firstpage}


\section{Introduction}
Hybrid Ermakov-Painlev\'e II systems were originally derived in \cite{cr14} via wave packet representations admitted by multi-dimensional coupled nonlinear Schr\"odinger systems incorporating de Broglie-Bohm potential terms. Therein, in particular, the canonical single component Ermakov-Painlev\'e II equation was shown to occur notably in the analysis of transverse wave propagation in a generalised Mooney-Rivlin hyperelastic material. It has subsequently been derived in such diverse physical applications as cold plasma physics \cite{crpc18}, Korteweg capillarity theory \cite{crpc17} and in the analysis of Dirichlet boundary value problems which arise out of the classical Nernst-Planck electrolytic system \cite{pacr15}. Underlying integrable structure in multi-component Ermakov-Painlev\'e II systems has been delimited in \cite{crws16} wherein an admitted Ermakov system invariant was applied to construct an algorithmic solution procedure involving a Painlev\'e II connection. A B\"acklund transformation admitted by Painlev\'e II was applied iteratively therein and the important link between the Ermakov-Painlev\'e II and integrable Painlev\'e XXXIV equation set down.

The application of the inverse scattering transform and its developments to solve initial/boundary problems in modern soliton theory has an extensive literature (qv \cite{mapc91,bk92,cr22} and work cited therein). However, the analysis of moving boundary problems of Stefan-type for solitonic equations has but a recent origin with motivation in the classical Saffman-Taylor model of \cite{psgt58}. The latter was concerned with description of the percolation of a liquid into a porous medium or Hele-Shaw cell. In a remarkable later development in \cite{gvlk91}, the canonical Dym equation of soliton theory \cite{pv01} was derived in a related analysis of the motion of the interface between a viscous and non-viscous liquid. In \cite{cr15}, a Painlev\'e II symmetry reduction was applied to derive exact solutions to a class of moving boundary problems of generalised Stefan-type for the Dym equation and its reciprocal associates. This type of reduction had its origin in an analysis of the evolution of the interface in a Hele-Shaw cell \cite{afst98}. In \cite{wscr99}, an extended S-integrable version of the Dym equation was derived in the geometric context of binormal motion of inextensible curves. In terms of physical application, this generalisation occurs in the hydrodynamics of unidirectional dispersive shallow water propagation with novel peaked soliton phenomena (Camassa and Holm \cite{rcdh99}). Iterated action of a classical B\"acklund transformation admitted by Painlev\'e II was applied to this extended Dym equation in \cite{cr17} to generate exact solution to a novel class of Stefan-type moving boundary problems in terms of Yablonski-Vorob'ev polynomials.

In modern soliton theory, reciprocal transformations associated with admitted conservation laws were introduced in \cite{jkcr84} and subsequently applied in \cite{crpw84} in the linkage of the canonical AKNS and WKI inverse scattering schemes of \cite{madkanhs73} and \cite{mwkkyi79} respectively. Reciprocal transformations likewise, importantly, connect certain classes of 1+1-dimensional solitonic hierarchies \cite{crws02}. In 2+1-dimensional systems, reciprocal-type transformations were constructed in \cite{cr85} and subsequently applied in soliton theory in the linkage of the triad of canonical Kadomtsev-Petviashvili, Dym and modified Kadomtsev-Petviashvili hierarchies \cite{wocr93}.

Moving boundary problems in 1+1-dimensional soliton theory of Stefan-type and their reciprocal associates which are amenable to exact solution via Painlev\'e II symmetry reduction have been detailed in \cite{cr16,cr23,cr25,cr2025}. In \cite{cr2022}, a reciprocal transformation allied with a M\"obius-type mapping was applied to a class of Stefan-type moving boundary problems for the solitonic Dym equation to generate exact parametric solution to a class for a base member of the WKI inverse scattering scheme. It is remarked that conjugation of reciprocal transformations and involutory-type transformations with origin in the autonomisation of Ermakov systems \cite{cacrurao90} has application in soliton theory.

A novel variant with temporal modulation of the modified Kadomtsev--Petviashvili equation of 2+1-dimensional soliton theory is presented here which admits Ermakov-Painlev\'e II symmetry reduction. The latter is applied to obtain exact solution to a class of nonlinear moving boundary problems of Stefan-type associated. A 2+1-dimensional generalisation of involutory-type transformations with origin in autonomisation of the canonical Ermakov-Ray-Reid system in \cite{cacrurao90} is applied to embed the extended mKP equation in a wide associated class with temporal modulation which inherits admittance of Ermakov-Painlev\'e II symmetry reduction.

\section{Ermakov-Painlev\'e II Symmetry Reduction} 
The canonical solitonic modified Kadomtsev-Petviashvili (mKP) equation is given by \cite{bk92}
\begin{equation} \label{1}
V_t+V_{xxx}-3V^2V_x-3V_x\partial^{-1}_xV_y+3\sigma^2\partial^{-1}_xV_{yy}=0\ , \quad \sigma^2=\pm1 \end{equation}
and, in particular, is amenable to the $\bar{\partial}$-dressing procedure of soliton theory \cite{bkvd92}. Here, a novel real variant of the preceding is introduced which admits Ermakov-Painlev\'e II symmetry reduction, namely
\begin{equation} \label{2}
\begin{array}{c} U_t+U_{xxx}-3U^2U_x-3U_x\partial^{-1}_xU_y+\delta^*U\partial^{-1}_xU_{yy} \\[2mm]
+\lambda(t+a)^\mu U^{-4}U_x=0. \end{array}\end{equation}
Here, an ansatz similarity reduction with
\begin{equation} \label{3}
U=(t+a)^m\Psi\left(\frac{x+\alpha^*y}{(t+a)^n}\right) \end{equation}
is postulated. Under the latter
\begin{equation} \label{4}
\partial^{-1}_xU_y=\alpha^*(t+a)^m\Psi=\alpha^*U\ , \ \partial^{-1}_xU_{yy}=\alpha^{*2}(t+a)^{m-n}\Psi'=\alpha^*U_y=\alpha^{*2}U_x \end{equation}
whence with $\delta^*=3/\alpha^*$, \eqref{2} reduces to
\begin{equation} \label{5}
U_t+U_{xxx}-3U^2U_x+\lambda(t+a)^\mu U^{-4}U_x=0. \end{equation}
The latter constitutes an extension of the solitonic mKdV equation in $U(x,y,t)$ wherein $y$ occurs implicitly.

Insertion of the similarity representation \eqref{3} into \eqref{5} yields
\begin{equation} \label{6}
\begin{array}{l} m(t+a)^{m-1}\Psi+(t+a)^{m-1}(-n\xi)\Psi'+(t+a)^{m-3n}\Psi{'''} \\[3mm]
\qquad -3(t+a)^{3m-n}\Psi^2\Psi' + \lambda(t+a)^{\mu-3m-n}\Psi^{-4}\Psi'=0, \end{array}\end{equation}
wherein $\Psi=\Psi(\xi)$, \ $\xi=(x+\alpha^*y)/(t+a)^n$.

Thus,
\begin{equation} \label{7}
\begin{array}{l} m\Psi-(n\xi\Psi)'+n\Psi+(t+a)^{1-3n}\Psi{'''}  -3(t+a)^{2m+1-n}\Psi^2\Psi' \\[3mm]
\qquad +\lambda(t+a)^{\mu-4m-n+1}\Psi^{-4}\Psi'=0 \end{array}\end{equation}
whence, with $m=-1/3$, \ $n=1/3$ and $\mu=-2$ there results
\begin{equation} \label{8}
\Psi{'''}-3\Psi^2\Psi'-(1/3)(\xi\Psi)'+\lambda\Psi^{-4}\Psi'=0. \end{equation}
On integration, the latter yields
\begin{equation} \label{9}
\Psi{''}-\Psi^3-(1/3)\xi\Psi-(\lambda/3)\Psi^{-3}=\zeta, \ \zeta\ \in \mathbb{R} \end{equation}
which, on appropriate scalings $\Psi=\gamma w^*$, and $\xi=\epsilon z$ and with $\zeta=0$ reduces to the canonical Ermakov-Painlev\'e II equation \cite{cr14} for $w^*(z)$:
\begin{equation} \label{10}
(w^*)''=2w^{*3}+zw^*+\chi w^{*-3} \end{equation}
with $\chi=(\lambda/3)\gamma^{-4}\epsilon^2$ upon choosing the scaling parameters so that
${e^3}/3=1$ and $\epsilon^2\gamma^2=2$, namely
$\epsilon= 3^{1/3}$, $\gamma= {2^{1/2}}/\epsilon$. 

\section{A Class of Stefan-Type Moving Boundary Problems}

Here, moving boundary problems associated with $U(x, y, t)$ are introduced governed by the nonlinear system
\begin{equation} \label{11}
\begin{array}{c} U_t+U_{xxx}-3U^2U_x+\lambda(t+a)^{-2}U^{-4}U_x=0 \\[4mm]
\left.\begin{array}{c} U_{xx}-U^3-(\lambda/3)(t+a)^{-2}U^{-3}=L_m S^i\dot{S} \\[2mm]
U=P_mS^j \end{array} \right\} \ \text{on} \ x+\alpha^*y=S(t), \ t>0 \\[7mm]
(U_{xx}-U^3-(\lambda/3)(t+a)^{-2}U^{-3})|_{x+\alpha^*y=0}=H_0(t+a)^k, \ t>0\ . \end{array}\end{equation}
wherein $S(t)=\gamma(t+a)^{1/3}$.

$$\textbf{Boundary Conditions}$$
I. $U_{xx}-U^3-(\lambda/3)(t+a)^{-2}U^{-3}=L_m S^i\dot{S}$  
{on  $x+\alpha^*y=S(t)=\gamma(t+a)^{1/3}, \, t>0$.}

\medskip

Insertion of the similarity representation \eqref{3} with $m=-1/3$, \ $n=1/3$ into the preceding yields
\begin{equation} \label{12}
\Psi{''}(\gamma)-\Psi^3(\gamma)-(\lambda/3)\Psi^{-3}(\gamma)=L_m(t+a)S^i\dot{S}. \end{equation}
On application of the Ermakov-Painlev\'e II reduction corresponding to $\zeta=0$ the relation
\begin{equation} \label{13}
L_m=\gamma\Psi(\gamma) \end{equation}
results together with $i=-1$.\\[3mm]
II. \ $U=P_mS^j$ \quad on \quad $x+\alpha^*y=S(t)=\gamma(t+a)^{1/3}, \quad t>0$.\\

Here, the relation
\begin{equation} \label{14}
P_m=\gamma\Psi(\gamma) \end{equation}
is derived along with $j=-1$.\\[3mm]
III. \ $(U_{xx}-U^3-(\lambda/3)(t+a)^{-2} U^{-3})|_{x+\alpha^*y=0}=H_0(t+a)^k.$\\

This yields
\begin{equation} \label{15}
\Psi{''}(0)-\Psi^3(0)-(\lambda/3)\Psi^{-3}(0)=H_0(t+a)^{k+1} \end{equation}
whence, by virtue of the Ermakov-Painlev\'e II reduction in $\Psi(\xi)$ at $\xi=0$ it is required that $H_0=0$.
 
\section{An Airy Reduction}

The Ermakov-Painlev\'e II equation \eqref{10} with $w^*=\rho^{1/2}$, \ $\rho>0$ on appropriate scaling yields
\begin{equation} \label{16}
\rho_{zz}=(\rho_z)^2/2\rho+2\rho^2+z\rho+2\delta/\rho\ , \quad \delta \in  \mathbb{R} \end{equation}
which is equivalent to the classical integrable Painlev\'e XXXIV equation (qv \cite{crpc17}, \cite{crws16}). In particular, it admits the solution
\begin{equation} \label{17}
\rho(z)=w_z+w^2+(1/2)z \end{equation}
wherein $w(z)$ is governed by the canonical Painlev\'e II equation \cite{crpc17}. Thus, the Ermakov-Painlev\'e II equation in turn admits a class of exact solutions
\begin{equation} \label{18}
\Psi=\gamma w^*=\gamma[\ w_z+w^2+(1/2)z\ ]^{1/2}. \end{equation}

Painlev\'e II admits an important subclass of exact solutions when the PII parameter $\alpha=1/2$, namely, that with
\begin{equation} \label{19}
w=-\phi'(z)/\phi(z) \end{equation}
with $\phi(z)$ governed by the classical Airy equation
\begin{equation} \label{20}
\phi{''}+(1/2)z\phi=0. \end{equation}
Accordingly \eqref{18} shows that the Ermakov-Painlev\'e II in $\Psi$ admits a particular class of solutions
\begin{equation} \label{21}
\Psi=\gamma[\ 2(\phi'/\phi)^2+z\ ]^{1/2}, \quad z=\xi/\epsilon \end{equation}
wherein
\begin{equation} \label{22}
\phi=aAi(2^{-1/3}z)+bBi(2^{-1/3}z), \quad a,  b \in \mathbb{R}.\end{equation}
Ermakov-Painlev\'e II solutions of the preceding Airy-type have physical application in \cite{crpc17} to the classical Korteweg capillarity system. Therein, the linked Painlev\'e XXXIV equation arises as a symmetry reduction of a Bernoulli integral of motion. It is remarked that in \cite{crbmkcha10}, the Airy-type solution \eqref{19} of Painlev\'e II with $\alpha=1/2$ and the subsequent class of exact solutions as generated via the iterated action of an admitted B\"acklund transformation has been applied in the analysis of certain boundary value problems for the Nernst-Planck electrolytic system.

In the present context of the class of moving boundary problems determined by the nonlinear system \eqref{11} with the preceding class of exact solutions of Airy-type for the Ermakov-Painlev\'e II equation, the parameters $L_m$ and $P_m$ in the moving boundary conditions are determined by
\begin{equation} \label{23}
L_m=P_m=\gamma\Psi(\gamma) \end{equation}
with
\begin{equation} \label{24}
\Psi(\gamma)=\gamma[\ 2(\phi'/\phi)^2+z\ ]^{1/2}|_{z=\gamma/\epsilon} \end{equation}
and $\phi$ determined in terms of $Ai(-z^{-1/3}z)$, \ $Bi(-2^{-1/3}z)$ by \eqref{22}.

It is remarked that the iterative action of the B\"acklund transformation admitted by Painlev\'e II as applied in the analysis of boundary problems for the Nernst-Planck system in \cite{crbmkcha10} may be used to generate, with \eqref{21} as seed solution, a wide class of associated solutions $\Psi$ of the Ermakov-Painlev\'e II equation and thereby of moving boundary problems of which the latter constitutes a symmetry reduction.

 \section{Temporal Modulation via a Class of Involutory Transformations}

Under the action on \eqref{5} of the class of transformations

\begin{equation} \label{25}\tag{$I^*$}
\begin{array}{c} \begin{array}{c} dt^*=\rho^{-2}(t)dt\ , \quad dx^*=dx\ , \quad dy^*=dy \\[3mm]
U^*=U/\rho(t)\ , \quad \rho^*=1/\rho(t) \end{array}  \end{array}\end{equation}
there results a diverse range of associated nonlinear evolution equations which incorporate temporal modulation, namely
\begin{equation} \label{26}
\begin{array}{c} \rho^{*2}\partial/\partial t^*(\rho^{*-1}U^*)+\rho^{*-1}U^*_{x^*x^*x^*}-3\rho^{*-3}U^{*2}U^*_{x^*} \\[3mm]
 + \lambda(t+a)^\mu\rho^{*3}U^{*-4}U^*_{x^*}=0 \end{array}\end{equation}
wherein $dt=\rho^{*-2}dt^*$. Under $I^*$,
\begin{equation} \label{27}
\begin{array}{c} dt^{**}=\rho^{*-2}dt^*=dt\ , \quad dx^{**}=dx\ , \quad dy^{**}=dy, \\[3mm]
U^{**}=U^*/\rho^*=U\ , \quad \rho^{**}=1/\rho^*=\rho, \end{array}\end{equation}
so that the involutory property $I^{**}=\mathrm{I}$ holds. Thereby, application of $I^*$ to the subclass of U determined by \eqref{3} which admits Ermakov-Painlev\'e II symmetry reduction generates a wide associated class with temporal modulation which inherits this property. Application of the involutory transformations $I^*$ to the moving boundary problems determined by the system \eqref{11} embeds them in a wide class which admits exact solution via Ermakov-Painlev\'e II symmetry reduction.

Involutory transformations $I^*$ of the type applied here have their genesis in a geometric analysis of coupled two-component Ermakov-Ray-Reid systems. The latter have application notably in nonlinear optics \cite{crbmkcha10,crbmha12} and 2+1-dimensional shallow water hydrodynamics \cite{crha10}. They may be set in the context of generalised nonlinear coupled Ermakov systems which have extensive applications in both physics and continuum mechanics \cite{crws18}.

Ermakov-type modulation of physical systems has been detailed in \cite{cr2014,crgsvv16,cr19,cr2023} notably in connection with nonlinear Schr\"odinger models and Kepler triads. In \cite{crwsbm20}, spatially modulated coupled systems of sine-Gordon, Demoulin and Manakov NLS type were reduced to their unmodulated solitonic counterparts via involutory transformations. Thereby canonical solitonic properties were inherited by the modulated systems.

In the present context, a wide range of temporal modulation of \eqref{5} is generated if $\rho^*(t^*)$ in the class of involutory transformations $I^*$ is governed by the classical Ermakov equation
\begin{equation} \label{28}
\rho^*_{t^*t^*}+\omega(t^*)\rho^*=\mathcal{E}/\rho^{*3}\ , \quad \mathcal{E}\ \in \mathbb{R}. \end{equation}
Application may then be made of its nonlinear superposition principle \cite{crws18}. The latter has physical application, in particular, in the analysis of initial-boundary value problems descriptive of the large amplitude radial oscillations of thin shells composed of hyperelastic materials of Mooney-Rivlin type and subject to a range of boundary loadings \cite{crwa89}. The nonlinear superposition principle may be derived via a Lie group procedure as in \cite{crur89}. Lie theoretical generalisation and discretisation of Ermakov-type equations which preserve admittance of nonlinear superposition principles have been detailed in \cite{crwspw97}.

\section{Conclusion}

Here, moving boundary problems of Stefan-type have been shown to be amenable to exact solution via Ermakov-Painlev\'e II symmetry reduction of a novel class of 2+1-dimensional nonlinear evolution equations. It is remarked that the reduction procedure and application of involutory transformations to incorporate temporal modulation may likewise be adapted to certain variants of the 2+1-dimensional Bogoyavlensky-Konopelchenko equation with its diverse physical applications.

\subsection*{Acknowledgements}

We thank the anonymous reviewer for the careful reading of the manuscript and his/her fruitful suggestions.

\label{lastpage}
\end{document}